\documentclass[11pt]{article}
\usepackage{graphicx,psfrag,amsmath}
\usepackage{fullpage}
\newcommand{\BEAS}{\begin{eqnarray*}}
\newcommand{\EEAS}{\end{eqnarray*}}
\newcommand{\BEQ}{\begin{equation}}
\newcommand{\EEQ}{\end{equation}}
\newcommand{\BIT}{\begin{itemize}}
\newcommand{\EIT}{\end{itemize}}

\newcommand{\eg}{{\it e.g.}}
\newcommand{\ie}{{\it i.e.}}

\newcommand{\reals}{{\mbox{\bf R}}}

\newcommand{\symm}{{\mbox{\bf S}}}  % symmetric matrices

\newcounter{oursection}

\usepackage{array}
\usepackage{url}
\usepackage{color}

\usepackage{caption,subcaption}

\newif\ifarxiv
\arxivtrue
\title{Optimal Current Waveforms for Switched-Reluctance Motors}

\author{
Nicholas Moehle\thanks{
Mechanical Engineering Department, Stanford University. 
\texttt{moehle@stanford.edu}}
\and Stephen Boyd\thanks{
Electrical Engineering Department, Stanford University. 
\texttt{boyd@stanford.edu}}
}
\date{\today}
\begin{document}
\maketitle

\begin{abstract}
In this paper, we address the problem of finding current waveforms
for a switched reluctance motor
that minimize a user-defined combination of torque ripple and RMS current.
The motor model we use is fairly general,
and includes magnetic saturation,
voltage and current limits,
and highly coupled magnetics
(and therefore, unconventional geometries and winding patterns).
We solve this problem by approximating it as a mixed-integer convex program,
which we solve globally using branch and bound.
We demonstrate our approach on an experimentally verified
model of a fully pitched switched reluctance motor,
for which we find the globally optimal waveforms, even for high rotor speeds.
\end{abstract}

\section{Introduction}
We consider the problem of choosing optimal current waveforms for a
switched reluctance motor (SRM).
Traditionally, the shape of the current waveforms
is determined in an ad-hoc manner
(\eg, by fixing the turn-on angles for each phase winding
as a function of rotor position;
see \cite{toliyat2012handbook}).
Although creating these waveforms does not require a detailed motor model,
and implementing them is simple,
waveforms produced in this manner rarely produce
smooth output torque,
and often decrease motor efficiency
and exacerbate mechanical vibration and acoustic noise issues.
Also, because of inverter voltage limits,
such waveforms may not even be realizable at high rotor speeds.

We therefore propose to use optimization to find current waveforms
that acheive a desired average torque, 
while minimizing a combination of resistive power loss and RMS torque ripple.
We consider supply voltage limits, 
as well as current limits in each phase winding.
Our model includes a detailed magnetic circuit,
which can account for magnetic coupling between phases,
and can be used to model motors with unconventional geometries and winding patterns,
such as those with fully pitched windings.

We propose to solve this optimization problem
by approximating it as a mixed-integer convex program (MICP).
This MICP reformulation approach has several inherent advantages
over more conventional methods, such as sequential quadratic programming.
The most prominent, for our purposes,
is that it can be solved globally by generic
methods such as branch and bound, often in a reasonable amount of time.
This has two benefits:
first, it allows us to achieve the best performance possible for a given motor;
second, it provides a metric against which other, suboptimal methods can be compared.
We note that although in general global optimization
methods for solving MICPs can have very high (exponential) runtime,
we find that global solutions can typically be found in a reasonable
amount of time (1-5 minutes) for the problems we encounter.
These waveforms can be computed and stored in a lookup table,
indexed by desired torque and rotor speed.
Such a table with hundreds or thousands of entries
could be computed overnight on a standard multi-core computer.

Another advantage of an MICP formulation is that,
if suboptimal solutions are acceptable,
first-order methods exist that can produce a good
solution very quickly 
especially when initialized with a decent initial guess.
(see, \ie, \cite{takapoui2015simple})
This opens the possibility that (nearly) optimal waveforms can
be produced by a smart control scheme
even as motor parameters vary over the life of the motor.
This is especially attractive when combined with a lookup table
containing precomputed, globally optimal waveforms:
the precomputed waveforms can be used as an initial guess
for an online optimization method
when new waveforms (corresponding to updated motor parameters) are required.

We demonstrate our approach numerically on the experimentally validated 
motor model of \cite{mecrow2001modeling},
which describes an SRM with fully pitched phases.

\subsection{Previous work}

\paragraph{Current optimization for SRMs.}
Several authors have considered optimization of SRM current waveforms.
The most similar work to our own is 
a series of papers by Lovatt and Stephenson 
\cite{lovatt1994computer},
\cite{lovatt1997computer},
\cite{lovatt1997optimum},
that use local optimization methods to find minimum RMS current waveforms
that acheive a given (average or pointwise) desired torque,
subject to voltage and current constraints.
%concentrated windings.
Stankovic et al.\
derive optimal waveforms for a simple SRM,
under several restrictive assumptions
(\eg, sufficient drive voltage, 
no more than two simultaneously conducting phases);
under these assumptions, it is only necessary to discretize the
waveforms during commutation.
% -using additional assumptions to get (quasi) analytical solutions
%   -exists a period with only one current nonzero
%   -assume only two currents nonzero otherwise
% -numerical opt is only over overlapping (commutating) periods
Kaiserseder et al.\ \cite{kaiserseder2003current}
also seek optimal current waveforms that produce smooth torque
output and can be chosen to either minimize current RMS values,
or minimize vibration resulting from radial forces.
The method of optimization is not described.
A similar optimization problem is posed by Chapman and Sudhoff 
\cite{chapman2002design}
in the frequency domain;
sequential quadratic programming is used to (approximately) solve it.
%-silly_Vsquared_objective_no_alg_given.pdf
%  -not so great, objective is squared phase velocity...
%  -might have some ok references
Many other works optimize over current waveforms parametrized by only
a few free variables, 
such as the firing angle or the corner locations of a trapezoidal waveform.
This of course requires predetermined current waveform shapes,
which are not optimal in general.
For some examples of this approach, see 
\cite{mademlis2003performance}, \cite{le1993current}, and \cite{choi2002new}.

We also note that our approach can be viewed as an extension
of the authors' previous work on optimal current waveform design
for permanent magnet motors; see \cite{moehle2015optimal}.

% -other not-so-good papers
%   -kjair
%     -ad hoc, analytical solutions to special case
%     -many such results in the literature
%     -we present general approach
% 
% NEED TO FIND:
% -finch, Switched reluctance motor excitation current
%   -need to find this paper...
% -look more in stephenson's work
%  \paragraph{Feedback linearization.}
% -do not account for voltage limits
% -not optimal?
% -see kaiserseder for refs

\paragraph{Hybrid control and MPC for SRMs.}
Here we list some other optimization-based techniques 
applied to control SRMs.
Peyrl et al.\ propose a finite-set model predictive control approach,
which involves online optimization directly over future inverter switching states.
Due to the computationally demanding nature of this technique,
only very short prediction horizons (\ie, up to three steps) are considered.
The work by Vasak et al.\ \cite{vasak2007bounding},
uses a piecewise-affine model of the torque characteristic 
(as we do)
to derive a feedback controller that guarantees a torque ripple.
However, their proposed model is relatively low fidelity:
the dynamics are modelled as a first-order linear system,
and the proposed torque characteristic 
(our $g_k(\mathcal F_k, \theta)$) has only nine regions
(For comparison, in \S\ref{s-example} we use over one hundred regions.)

% \paragraph{Modeling of SRMs}
% Many aspects of our SRM model structure is from previous work.
% -fully pitched windings
% -magnetic circuits, graph theory
% -harmonic analysis

\paragraph{MICP.}
Convex optimization problems can be solved efficiently and reliably using standard techniques 
\cite{boyd2004convex}
(and additionally, specialized modelling software,
such as CVX \cite{cvx},  % and CVXPY [cvxpy], % XXX
enables rapid development of convex optimization applications).
An optimization problem with some integer variables,
but which is otherwise convex, is called a mixed integer convex program (MICP).
Because of the presence of integer variables,
MICPs are nonconvex optimization problems, 
and are difficult to solve (globally) in general
(\ie, these problems are are NP-hard; see
\cite{kleinberg2006algorithm}).
Indeed, all known global solution techniques for MICPs
(such as the branch-and-bound algorithm),
have exponential worst-case runtime.
Nevertheless, many of these algorithms are effective in practice,
and we found them to work well 
for the the optimization problem we formulate in this paper.
In addition to global solution algorithms,
many approaches exist to (approximately) solve MICPs
more quickly;
see \cite{takapoui2015simple}  % and [takapoui-diamond], % XXX
and references therein.

Our formulation is based on approximating the nonlinear constraint functions
(the torque characteristics and the magnetic flux characteristics)
by piecewise affine functions.
These constraints can then be represented as a combination
of integer and linear constraints using disjunctive programming.
For details on disjunctive programming consult
Balas \cite{balas1979disjunctive} and Ceria and Soares \cite{ceria1999convex}.
Disjunctive programming has found many applications in the past decade or so,
such as process engineering \cite{grossmann2013systematic},
facility location, unit commitment and portfolio management \cite{gunluk2012perspective},
and optimal control \cite{moehle2015perspective}.

%Our approach to handling the nonlinear constraints is one of many possible techniques.
%Another very effective technique worth mentioning is based the convex-concave procedure;
%see [cvx-ccv] % XXX

\subsection{Contribution}
Our reformulation of the torque control problem as a MICP opens the door for two
interesting possiblilites.
The first is that global solution methods can be used to find the optimal waveforms,
typically in a reasonable amount of time.
This is of course advantageous in its own right,
as it allows provably optimal waveforms to be implemented.
It is also useful to verify the limits of performance of motors,
and as a benchmark for comparing heuristic methods.
The second possibility is that fast first-order methods, 
which are simple enough to run on embedded platforms,
can be used as a heuristic to update the globally optimal waveforms as 
parameters vary over the life of the motor.
Our proposed model is also much more general than
the optimization models considered in previous works,
and can therefore be used to capture more of the characteristc features of
switched reluctance motors, such as magnetic coupling.
We hope this generality will be useful for researchers investigating
novel switched reluctance motor topologies,
by giving them a practical method for optimal waveform generation,
and by characterizing the theoretical performance of their designs.

\section{Motor model}
\label{s-model}
We consider an abstract, lumped parameter model of a switched reluctance motor.
The rotor, which does not contain any windings or magnetic elements,
has angular position $\theta$ and angular velocity $\omega$;
we assume $\omega$ is constant. 
The stator contains $n$ electrical circuit branches, called {\it windings}.
The winding currents are $i \in \reals^n$,
the winding voltages are $v \in \reals^n$,
and the magnetic flux linkages through the windings are $\lambda \in \reals^n$.
The stator also contains $m$ magnetic elements,
with magnetomotive force (MMF) vector $\mathcal F \in \reals^m$,
and magnetic flux vector $\psi \in \reals^m$.

We will assume that 
$i$, $v$, $\lambda$, $\mathcal F$, and $\psi$
are $2\pi$-periodic functions of $\theta$.
We use a prime ($'$) to denote differentiation of these functions with respect
to $\theta$. 
To lighten notation, we often drop explicit dependence on $\theta$.
%Multiplication and addition is taken to be pointwise.

\paragraph{Electrical dynamics.}
The electrical circuit dynamics are
\begin{equation}
\label{e-dynamics}
v = Ri + \omega \lambda',
\end{equation}
where $R\in\symm_{++}^n$ is the (diagonal) resistance matrix.
Note that $\omega\lambda'$ is the time derivative of $\lambda$.

\paragraph{Magnetic circuit.}
We assume the magnetic elements are connected by a (planar) magnetic circuit,
which we describe in terms of mesh analysis
%a common method for analyzing electrical and magnetic circuits.
(for an introduction to mesh analysis, see \cite{desoer1984basic}).
%We assume this magnetic circuit is planar,
%although with a more sophisticated approach (\eg, loop analysis)
%the assumption can be removed.
In particular, we assume that there are $l$ circuit meshes
(not including the outer mesh)
each with an associated reference direction.
The (reduced) \emph{mesh matrix} $M\in \reals^{l \times m}$
is such that 
% $M_{jk} = 1$ if mesh $j$ includes magnetic element $k$,
% and their reference directions agree, 
% $M_{jk} = -1$ if they do not agree, and 
% $M_{jk} = 0$ otherwise.
\[
M_{jk} = 
\begin{cases}
 1 & \text{if magnetic element $k$ is in mesh $j$, with coinciding reference directions} \\
-1 & \text{if magnetic element $k$ is in mesh $j$, with opposite reference directions} \\
 0 & \text{if magnetic element $k$ is not in mesh $j$}.
\end{cases}
\]
Any flux vector $\psi$ consistent with the magnetic circuit topology
must be a linear combination of the rows of $M$, so that
\begin{equation}
\label{e-loop-kcl}
\psi = M^T \phi,
\end{equation}
for some $\phi(\theta) \in \reals^l$,
which we call the \emph{mesh magnetic flux vector}.
(Because $M$ has full row rank in general, 
there is a unique $\phi$ for any such $\psi$.)

In addition, the total MMF around each mesh 
is the sum of the MMFs of the magnetic elements
that make up the mesh (accounting for reference direction),
so the vector of mesh MMFs is given by the vector $M \mathcal F$.

\paragraph{Electro-magnetic geometry.}
We define the \emph{electro-magnetic geometry matrix}
$C \in \reals^{l \times n}$
such that $C_{jk}$ gives the amount of current passing through
mesh $j$ per unit of current in winding $k$.
The total current passing through each of the $l$ meshes is therefore
given by the vector $Ci$,
which is related to the total MMF around the meshes by Amp\`ere's law:
\begin{equation}
\label{e-faraday}
M \mathcal F = Ci.
\end{equation}
Similarly, the flux linkage is related to the mesh magnetic flux vector by
\begin{equation}
\label{e-flux-linkage}
\lambda = C^T \phi.
\end{equation}

\paragraph{Magnetic characteristic.}
The magnetic flux and the MMF of the $k$-th magnetic element are related by
\begin{equation}
\label{e-magnetic-characteristic}
\psi_k(\theta) = f_k\big(\mathcal F_k(\theta), \theta\big),
\end{equation}
where the magnetic characteristic $f_k$ 
is a monotonically increasing function in its first argument. 

As a special case, if the functions $f_k$ are affine in $\mathcal F_k$
for each $\theta$, with constant linear term
(so that $\psi(\theta) = A \mathcal F(\theta) + b(\theta)$,
with $A$ diagonal and positive definite),
as in the case of a permanent magnet motor,
then 
%using (\ref{e-mmf-current}) and (\ref{e-phi-psi}),
we have $\lambda(\theta) = Li(\theta) + k(\theta)$,
where 
$L = C^T (MA^{-1}M^T)^{-1}C$
is the \emph{inductance matrix} and 
$k(\theta) = C^T (MA^{-1}M^T)^{-1}A^{-1} Mb(\theta)$
is the \emph{back-emf constant}. 

% \paragraph{\emph{proof}.}
% We have 
% \[
% M^T\phi = A \mathcal F + b.
% \]
% Then we can write
% \[
% \begin{bmatrix}
% -A & M^T \\
% M & 0
% \end{bmatrix}
% \begin{bmatrix}
% \mathcal F \\ \phi
% \end{bmatrix}
% =
% \begin{bmatrix}
% b \\ Ci
% \end{bmatrix}.
% \]
% Solving for $\mathcal F$ gives
% \[
% A^{-1}M^T\phi - A^{-1}b = \mathcal F.
% \]
% Then we have
% \begin{align*}
% Ci 
% &= M\mathcal F \\
% &= MA^{-1}M^T\phi - A^{-1}Mb \\
% \end{align*}
% This gives
% \begin{align*}
% \phi 
% &= (MA^{-1}M^T)^{-1} (Ci + A^{-1}Mb) \\
% &= (MA^{-1}M^T)^{-1} Ci + (MA^{-1}M^T)^{-1} A^{-1}Mb
% \end{align*}
% We then have
% \begin{align*}
% \lambda 
% &= C^T \phi \\
% &= C^T(MA^{-1}M^T)^{-1} Ci + C^T(MA^{-1}M^T)^{-1} A^{-1}Mb).
% \end{align*}

\paragraph{Torque.}
The magnetic co-energy is
\[
E^*(\mathcal F, \theta) 
  %= \sum_{k=1}^m \int_0^{\mathcal F_k} \psi_k \; dx
  = \sum_{k=1}^m \int_0^{\mathcal F_k} f_k(x, \theta) \; dx.
\]
The electromagnetic torque can be expressed as
\begin{equation*}
%\label{e-torque}
\tau(\theta) 
  = -\frac{\partial}{\partial \theta} E^*\big(\mathcal F(\theta), \theta \big)
  = -\sum_{k=1}^m \frac{\partial}{\partial \theta}
    \int_0^{\mathcal F_k} f_k(x, \theta) \; dx.
\end{equation*}
By introducing a \emph{phase torque} function $g_k$ such that 
\[
g_k(y, \theta)
  = -\frac{\partial}{\partial \theta}
    \int_0^y f_k(x, \theta) \; dx,
\]
we have 
\begin{equation}
\label{e-sum-phase-torque}
\tau(\theta) = \sum_{k=1}^m g_k \big(\mathcal F_k(\theta), \theta \big).
\end{equation}

\paragraph{Voltage limits.}
We assume the winding voltages are bounded:
\begin{equation}
\label{e-voltage-limit}
| v_k(\theta) | \leq v^{\rm max}, \quad k = 1, \ldots, n.
\end{equation}

\paragraph{Torque ripple.}
The average torque over one cycle is 
\[
\overline \tau = \frac1{2\pi } 
\int_0^{2\pi} \tau(\theta) \; d\theta .
\]
The (quadratic) torque ripple is
\[
r = \frac1{2\pi} 
\int_0^{2\pi} \big(\tau(\theta) -\overline \tau \big)^2 \; d\theta.
\]

\paragraph{Power loss.}
The power loss is the average resistive loss from the phase
currents over one cycle:
\[
P_{\rm loss} = \frac1{2\pi}
\int_0^{2\pi} 
i(\theta)^T R i(\theta) \; d\theta .
\]

\section{Optimal torque control}
The optimal torque control problem is to 
choose the phase voltages, phase currents, and eddy currents 
to achieve a desired average
torque while minimizing the average power loss and torque ripple:
\begin{equation}
\begin{array}{ll}
\mbox{minimize} & P_{\rm loss} + \alpha  r \\ %+ \sum_{l=1}^\infty \beta_l |\hat f_l|^2 \\
\mbox{subject to} & \overline \tau = \tau^{\rm des},  \\
                  & \mbox{equations
                    (\ref{e-dynamics}),
                    (\ref{e-loop-kcl}),
                    (\ref{e-faraday}),
                    (\ref{e-flux-linkage}),
                    %(\ref{e-mmf-current}),
                    %(\ref{e-phi-psi}),
                    (\ref{e-magnetic-characteristic}),
                    %(\ref{e-torque}),
                    (\ref{e-sum-phase-torque}),
                    %(\ref{e-vibration-dynamics}),
                    and
                    (\ref{e-voltage-limit}).
                    } \\
\end{array}
\label{e-opt-torque-ctrl}
\end{equation}
The parameters are the trade-off parameter $\alpha\geq 0$,
the rotor angular velocity $\omega$, 
the desired average torque $\tau^{\rm des}$,
the resistance matrix $R$,
the mesh matrix $M$,
the electro-magnetic geometry matrix $C$,
and the magnetic characteristic functions $f_k$, for $k = 1, \ldots, m$.
The problem variables are the $2\pi$-periodic functions
$i$, $v$, $\lambda$, $\mathcal F$, $\psi$, and $\phi$.

Problem (\ref{e-opt-torque-ctrl})
is an infinite-dimensional optimization problem.
The problem is nonconvex
due to the magnetic characteristic (\ref{e-magnetic-characteristic})
and the torque relation (\ref{e-sum-phase-torque}),
and therefore is expected to be difficult to solve globally.

One approach is to find a locally optimal solution,
using common local optimization methods such as sequential quadratic programming
Software that implements these methods is readily available;
see \cite{nocedal2006numerical}.
%(We suspect that for our problem, off-the-shelf methods
%such as Matlab's \texttt{fmincon} will work 

In this paper, we pursue a different approach,
and will instead show how to approach (\ref{e-opt-torque-ctrl})
by discretizing the variables,
and (approximately) reformulating the problem as a mixed-integer convex program (MICP),
a problem class for which efficient algorithms are available
to find a good (or even globally optimal) solution.
%Note that many algorithms are truly global algorithms,
%in that if they are run to completion, they are guaranteed to find the global solution.

We note that if the magnetic characteristic is affine,
with only the offset depending on rotor position,
the torque control problem reduces to a version of the formulation given in 
\cite{moehle2015optimal}.
Our current problem can therefore be interpreted as an extension
of that formulation to cover magnetic nonlinearities and reluctance torque.

\section{Conversion to MICP}
\label{s-micp}
In this section we show how to convert (\ref{e-opt-torque-ctrl}) to an
(infinite-dimensional) mixed-integer convex program.
To do this,
we use piecewise-affine approximations of the nonlinear equality constraints.

\subsection{Approximation of magnetic characteristic}
Here we approximate the equation magnetic characteristic
(\ref{e-magnetic-characteristic})
by a set of linear and integer constraints.
We replace the constraint 
$\psi_k = f_k(\mathcal F_k, \theta)$
with the constraint
\begin{equation}
\label{e-approx-magnetic-characteristic}
\psi_k = \tilde f_k(\mathcal F_k, \theta),
\end{equation}
where $\tilde f_k$ is a piecewise affine approximation of $f_k$.
In particular, we have
\[
\tilde f_k(x, \theta) = 
\begin{cases}
a_k^{1}(\theta) x + b_k^{1}(\theta) &
\tilde {\mathcal F}_k^{0}  \leq x \leq \tilde {\mathcal F}_k^{1} \\
\hfil \vdots  & \hfil \vdots\\
a_k^{N}(\theta) x + b_k^{N}(\theta) &
\tilde {\mathcal F}_k^{N-1}  \leq x \leq \tilde {\mathcal F}_k^{N},
\end{cases}
\]
where 
$a_k^1(\theta), \ldots, a_k^N(\theta)$
and
$b_k^1(\theta), \ldots, b_k^N(\theta)$
parametrize the affine functions, and 
$\tilde {\mathcal F}_k^0, \ldots, \tilde {\mathcal F}_k^N$
are the boundaries of the affine regions, so that we have
\[
f_k(x, \theta) \approx a_k^j(\theta) x + b_k^j(\theta)
\]
if $\tilde {\mathcal F}_k^{j-1} \leq x \leq \tilde {\mathcal F}_k^j$.

By introducing additional variables $z_k^j(\theta)$ and $s_k^j(\theta)$,
for $j = 1, \ldots, N$, and $k = 1, \ldots, m$,
the approximate magnetic characteristic constraint (\ref{e-approx-magnetic-characteristic})
can be written as 
\begin{equation}
\label{e-ftilde-constraints}
\begin{gathered}
\begin{aligned}
\psi_k(\theta) = \sum_{j=1}^N a_k^j(\theta) z_k^j(\theta) + b_k^j(\theta) s_k^j(\theta) 
&\qquad&
\mathcal F_k(\theta) = \sum_{j=1}^N z_k^j(\theta)
&\qquad&
\sum_{j=1}^N s_k^j = 1
\end{aligned}
\\
\begin{aligned}
\tilde {\mathcal F}_k^{j-1} s_k^j(\theta) \leq z_k^j(\theta) 
\leq \tilde {\mathcal F}_k^{j} s_k^j(\theta)
&\qquad&
s_k^j(\theta) \in \{0, 1\}.
\end{aligned}
\\
\end{gathered}
\end{equation}

\subsection{Approximation of torque function}
In the same way, we can approximate the torque constraint
(\ref{e-sum-phase-torque})
using a set of linear and integer constraints.
To do this, we first approximate 
each torque function $g_k$, for $k = 1, \ldots, m$,
as a piecewise affine function $\tilde g_k$:
\[
\tilde g_k(x, \theta) = 
\begin{cases}
c_k^{1}(\theta) x + d_k^{1}(\theta) &
\tilde {\mathcal F}_k^{0}  \leq x \leq \tilde {\mathcal F}_k^{1} \\
\hfil \vdots  & \hfil \vdots\\
c_k^{N}(\theta) x + d_k^{N}(\theta) &
\tilde {\mathcal F}_k^{N-1}  \leq x \leq \tilde {\mathcal F}_k^{N},
\end{cases}
\]
where 
$c_k^1(\theta), \ldots, c_k^N(\theta)$
and
$d_k^1(\theta), \ldots, d_k^N(\theta)$
parametrize the affine functions.
Then the approximate torque constraint 
\[
\tau(\theta) = \sum_{k=1}^m \tilde g_k(\mathcal F_k, \theta)
\]
can be included by appending
\begin{align}
\label{e-gtilde-constraints}
\tau(\theta) = \sum_{k=1}^m \sum_{j=1}^N  
   c_k^j(\theta) z_k^j(\theta) + d_k^j(\theta) s_k^j(\theta)
\end{align}
to the constraints (\ref{e-ftilde-constraints}) above.

\subsection{Improving the MICP formulation}
By converting the nonlinear constraints to linear and integer constraints,
we have succeeded in our goal of making (\ref{e-opt-torque-ctrl})
into an MICP.
However, the runtime of a global MICP solver often depends crucially on the problem formulation.
Here we give an additional reformulation of the objective of 
(\ref{e-opt-torque-ctrl})
(specifically, of the the power loss)
that may improve the performance of a MICP solver
compared with the basic formulation.

% After replacing the nonconvex functions with their integer-convex approximations,
% an MICP solver can be applied directly to solve the reformulated problem.
% However, because most MICP solvers use convex relaxations of the MICP,
% the tightness of the convex relaxation is crucial to the
% speed of the algorithm.  
% In this section we describe to reformulate the objective of ()
% to obtain a tighter convex relaxation.

Assuming that $C$ has full column rank, 
we can use (\ref{e-faraday}) to express the winding current as 
$i = C^\dagger M \mathcal F$,
where $C^\dagger$ is the pseudo-inverse of $C$
(or any other left inverse).
Then the power loss can be rewriten as
\[
P_{\rm loss} = \frac1{2\pi}
\int_0^{2\pi} 
\mathcal F(\theta)^T Q \mathcal F (\theta) \; d\theta,
\]
where $Q = M^T (C^\dagger)^T R C^\dagger M$.

For any feasible set of variables,
and for any diagonal matrix $D$, this is equivalent to
\[
\frac1{2\pi} \int_0^{2\pi} 
\Bigg(
\mathcal F(\theta)^T (Q-D) \mathcal F (\theta) 
%+ \sum_{k=1}^m \sum_{j=1}^N D_{kk} z_k^j(\theta)^2 / s_k^j(\theta)
+ \sum_{k=1}^m \sum_{j=1}^N D_{kk} \frac{z_k^j(\theta)^2}{s_k^j(\theta)}
\Bigg)
\; d\theta.
\]
In particular, if we choose $D$ with nonnegative diagonal elements,
such that $Q-D$ is positive semidefinite,
then the reformulated power loss function is convex in all its variables.
Furthermore, the larger the elements of $D$ are,
the tighter the convex relaxation will be.
Finding a suitable $D$ can be cast as a small convex optimization problem;
for details on this type of reformulation, see \cite{frangioni2007sdp}.

\section{Symmetry and discretization}

\subsection{Symmetry}
\label{s-symmetry}
Many motors have substantial symmetry,
which in our formulation is encoded in the functions $f_k$, 
as well as $M$, $R$, $C$.
Although it is \emph{not} necessarily true that 
optimal variables for (\ref{e-opt-torque-ctrl}) share this symmetry,
it is reasonable to expect that many good sets of variables are symmetric.
Furthermore, we may explicitly desire that the waveforms be symmetric
(\eg, to simplify implementation, or to wear components evenly).
By introducing symmetry constraints, 
we can also reduce the interval of the variables of (\ref{e-opt-torque-ctrl})
thus reducing the complexity of a 
discretized version of the problem.
For some asymmetric motors,
one or more of these assumptions may not hold;
examples of this include motors intentionally designed without symmetry,
or when a winding in an otherwise symmetric motor has failed.

\paragraph{Pole symmetry.}
We assume the rotor has $N_{\rm p}$ pole pairs
\ie, $f_k$ is $2\pi/N_{\rm p}$-periodic for all $k$.
Consequently, we restrict our search to
variables $i$, $v$, $\lambda$, $\psi$, $\phi$, and $\mathcal F$ that are also
$2\pi/N_{\rm p}$-periodic.

\paragraph{Phase symmetry.} 
We assume the motor has $K$ phases,
and therefore search for variables that satisfy the periodicity property
\[
\label{e-phase-symm-current-periodicity}
i_k(\theta) = i_1 \left(\theta + \frac{2\pi (k-1)}{K N_{\rm p}} \right),
\]
with similary constraints holding for
$v$, $\lambda$, $\mathcal F$, $\psi$, and $\phi$.

\paragraph{Equivalent problem.} 
The symmetry assumptions allow us to form an equivalent problem with 
the same constraints and objective as 
(\ref{e-opt-torque-ctrl})
in which the variables have domain 
$[0,2\pi/(KN_{\rm p})]$.  
We also add periodicity constraints of the form
\[
i_k(0) = i_1 \left(\frac{2\pi (k-1)}{K N_{\rm p}} \right),
\qquad \text{for $k = 1, \ldots K$}.
\]
with similar constraints for the other variables.
The integrands in the definitions of
average torque, torque ripple, and power loss 
are each $\pi/(3N_{\rm p})$-periodic; 
to get equivalent definitions of these values over the appropriate domain,
we can integrate over $[0,2\pi/(KN_{\rm p})]$ instead of $[0,2\pi]$, 
and scale the result by $KN_{\rm p}$.

\subsection{Discretization}
After reducing the domain of the variables of (\ref{e-opt-torque-ctrl}),
we discretize this interval
into $T+1$ grid points, 
$\theta_0,\ldots,\theta_T$, 
with $\theta_0=0$ and $\theta_T = 2\pi/(KN_{\rm p})$.  All
pointwise constraints must hold at $\theta_0,\ldots,\theta_{T-1}$, and
the periodicity constraints 
must hold at $\theta_0$ and $\theta_T$. 
Integration over the interval is replaced by summation from
$\theta_0$ to $\theta_{T-1}$, with appropriate scaling.   
We approximate the derivative in (\ref{e-dynamics}) 
using a forward difference approximation.

\section{Example}
\label{s-example}

\subsection{Motor model}
\begin{figure} 
\begin{center}
\psfrag{a}[][cc]{$\rm A$} 
\psfrag{b}[][cc]{$\rm B'$} 
\psfrag{c}[][cc]{$\rm C$} 
\psfrag{d}[][cc]{$\rm A'$} 
\psfrag{e}[][cc]{$\rm B$} 
\psfrag{f}[][cc]{$\rm C'$} 
\psfrag{1}{\color{white} $3'$} 
\psfrag{2}{\color{white} $1$} 
\psfrag{3}{\color{white} $2'$} 
\psfrag{4}{\color{white} $3$} 
\psfrag{5}{\color{white} $1'$} 
\psfrag{6}{\color{white} $2$} 
\ifarxiv
\includegraphics[width=.3\textwidth]{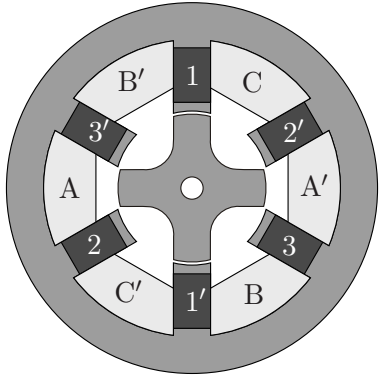}
\else
\includegraphics[width=.3\textwidth]{fig/sr_motor.eps}
\fi
\caption{
A switched reluctance motor with fully pitched windings.
Phase windings are shown in light gray, and are labeled A, B, and C
for current flowing out of the page
(with complementary flow denoted with a prime).
Magnetic elements are shown in dark gray, 
and are labeled 1, 2, and 3, with magnetic flow toward the rotor
(with complementary flow denoted with a prime).
Note that every third stator teeth belongs to a single magnetic element.
}
\label{f-sr-motor}
\end{center}
\end{figure}

We consider the switched reluctance motor with fully pitched windings shown in 
figure \ref{f-sr-motor}.
Our model is based the first example of \cite{mecrow2001modeling}.
The motor model has fully pitched windings,
instead of the more usual concentrated windings.
(This is reflected by the matrix $M$,
which would be diagonal for a motor with concentrated windings).

\paragraph{Symmetry and discretization.}
We assume that each stator tooth is magnetically identical,
which allows us to group every third stator tooth into a single magnetic element.
With this assumption,
the motor exhibits pole symmetry, with $N_{\rm p} = 2$,
and phase symmetry, with $K = 3$ phases.
We discretize the interval $[0, \pi/3]$ with $T = 40$,
so that $\theta_0 = 0$ and $\theta_T = \pi/3$.

\paragraph{Electromagnetic.}
We have $m = 3$ magnetic elements, and $l = 3$ meshes.
The mesh matrix of the magnetic circuit is
\[
M = 
\begin{bmatrix}
   0 &  1 &  1 \\
   1 &  0 &  1 \\
   1 &  1 &  0
\end{bmatrix}.
\]
The motor has $n = 3$ windings.
The voltage limit is $600 \; \rm V$.
%and has resistance matrix $R = r_{\rm winding} I$,
%where $r_{\rm winding} = 0.1 \Omega$ is the winding resistance,
%and $I$ is the ($3 \times 3$) identity matrix.
The resistance and electro-magnetic geometry matrices are
\[
R = 
\begin{bmatrix}
  0.1 &   0 &   0 \\
    0 & 0.1 &   0 \\
    0 &   0 & 0.1
\end{bmatrix} \; \Omega,
\qquad
C = (1/2)N_{\rm turns}
\begin{bmatrix}
  1 & 0 & 0 \\
  0 & 1 & 0 \\
  0 & 0 & 1
\end{bmatrix},
\]
where $N_{\rm turns} = 204$ is the number of turns in each winding.

% \paragraph{\emph{proof}.}
% Take all matrices to be invertible and symmetric.
% In Mecrow, we have 
% \[
% \lambda = N_{\rm phase} 
% \begin{bmatrix}
%   -1 &  1 &  1 \\
%    1 & -1 &  1 \\
%    1 &  1 & -1
% \end{bmatrix}
% \psi.
% \]
% Because $\lambda = C \phi$, $\psi = M \phi$, we have 
% $\lambda = C M^{-1} \psi$.
% So we must have
% \[
% C M^{-1}
% =
% N_{\rm phase}
% \begin{bmatrix}
%   -1 &  1 &  1 \\
%    1 & -1 &  1 \\
%    1 &  1 & -1
% \end{bmatrix}.
% \]
% They also have 
% \[
% \frac{n_{\rm stator}}{6N_{\rm phase}}
% \begin{bmatrix}
%    0 &  1 &  1 \\
%    1 &  0 &  1 \\
%    1 &  1 &  0
% \end{bmatrix}
% \mathcal F_{\rm mecrow}
% = i.
% \]
% Because $M\mathcal F = Ci$, 
% and $\mathcal F =\alpha \mathcal F_{\rm mecrow}$,
% this means that
% \[
% C^{-1}M\mathcal F = i,
% \]
% so that 
% \[
% \frac{n_{\rm stator}}{3N_{\rm phase}}
% \frac{1}{2}
% \begin{bmatrix}
%    0 &  1 &  1 \\
%    1 &  0 &  1 \\
%    1 &  1 &  0
% \end{bmatrix}
% = \alpha C^{-1}M.
% \]
% Because $n_{\rm stator} = 12$, Then we have
% \[
% \frac{3N_{\rm phase}}{n_{\rm stator}}
% \begin{bmatrix}
%   -1 &  1 &  1 \\
%    1 & -1 &  1 \\
%    1 &  1 & -1
% \end{bmatrix}
% =
% \frac{N_{\rm phase}}{4}
% \begin{bmatrix}
%   -1 &  1 &  1 \\
%    1 & -1 &  1 \\
%    1 &  1 & -1
% \end{bmatrix}
% = \alpha M^{-1}C.
% \]
% So we have $\alpha = 4$.

\begin{figure} 
\begin{center}
%\psfrag{theta}[cc]{$\theta$} 
\ifarxiv
\includegraphics[width=.4\textwidth]{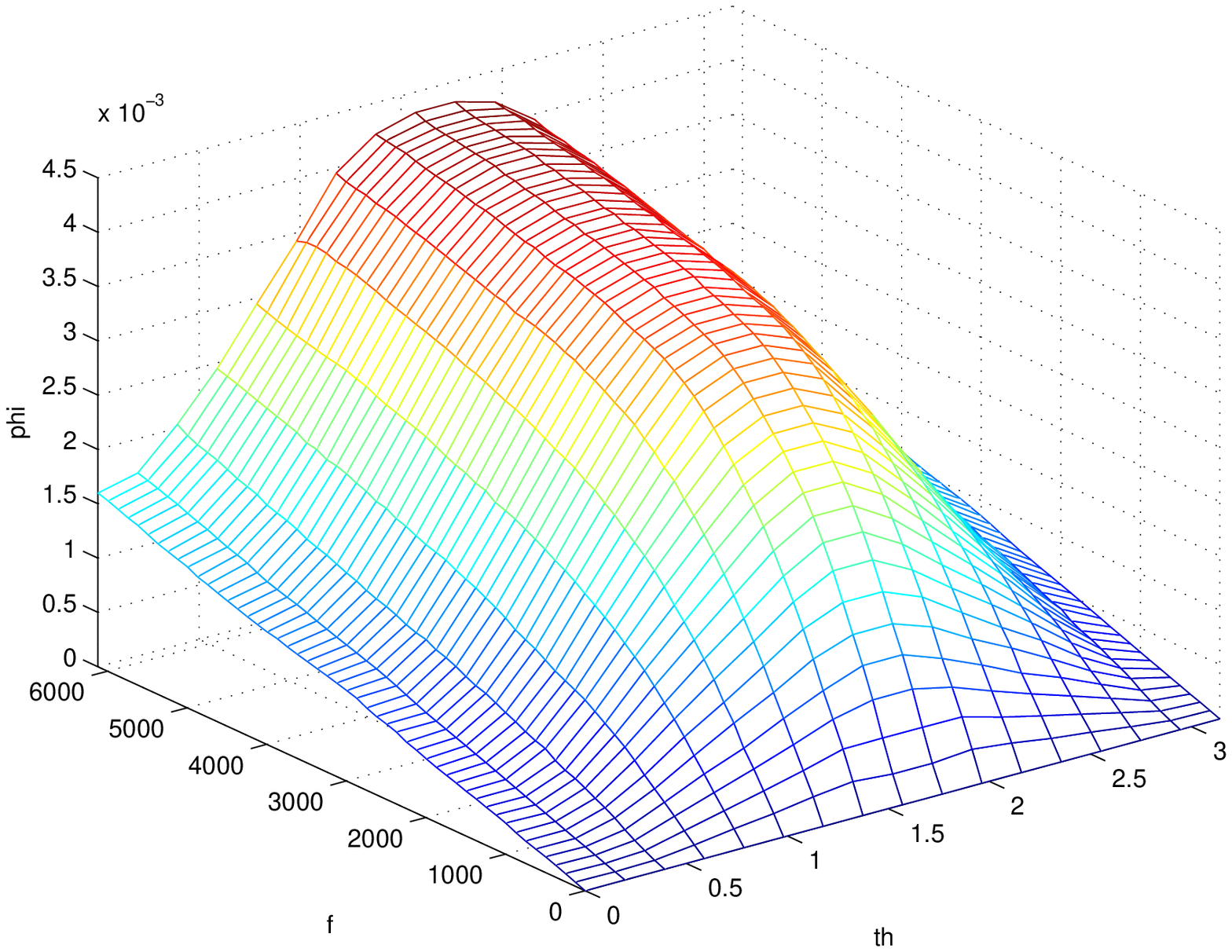}
\includegraphics[width=.4\textwidth]{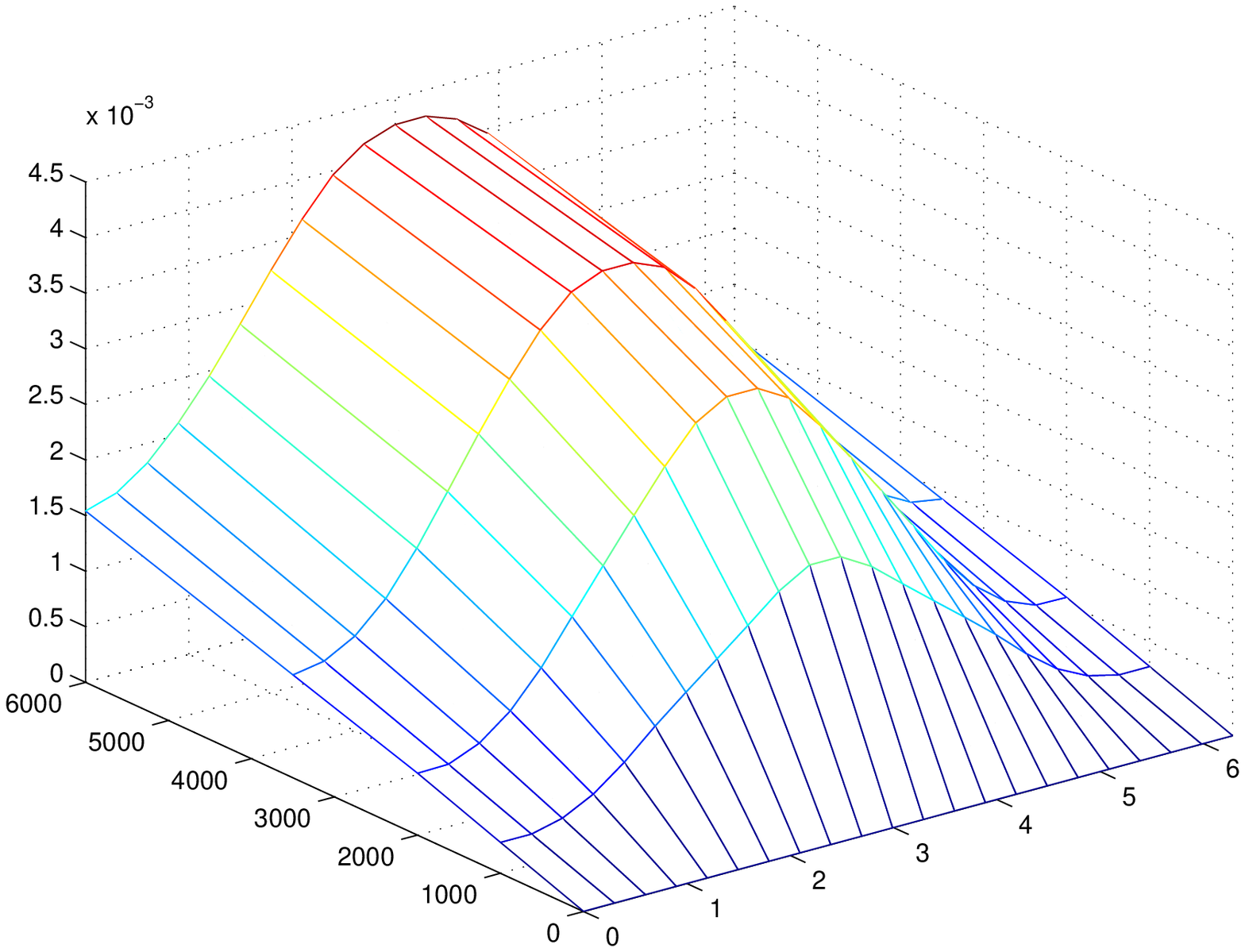}
\else
\includegraphics[width=.4\textwidth]{matlab/mecrow_model/psi.eps}
\includegraphics[width=.4\textwidth]{matlab/phi_hat.eps}
\fi
\caption{
The flux characteristic $f_1$ (left),
and its piecewise affine approximation $\tilde f_1$ (right).
}
\label{f-flux-characteristic}
\end{center}
\end{figure}

\begin{figure} 
\begin{center}
%\psfrag{theta}[cc]{$\theta$} 
\ifarxiv
\includegraphics[width=.4\textwidth]{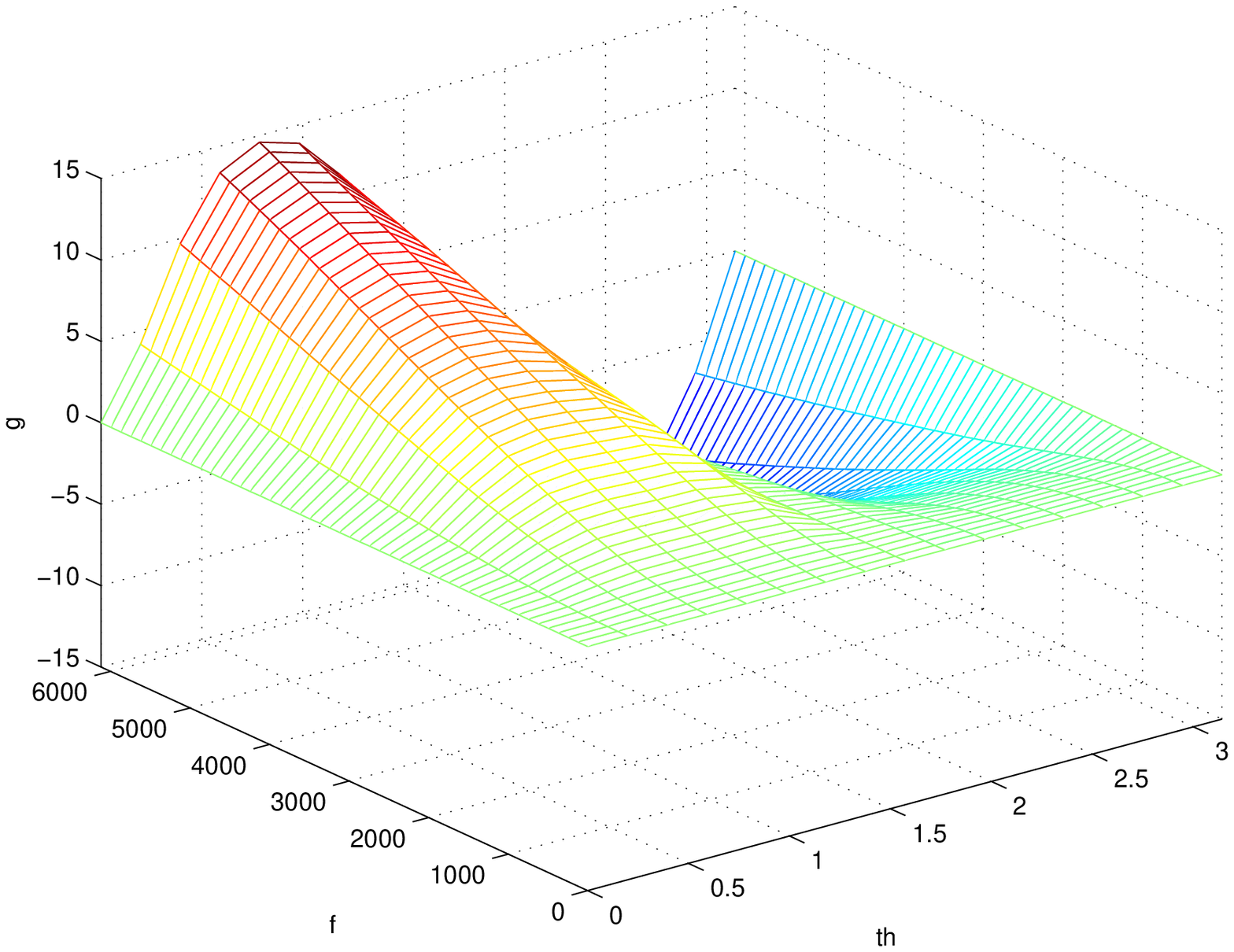}
\includegraphics[width=.4\textwidth]{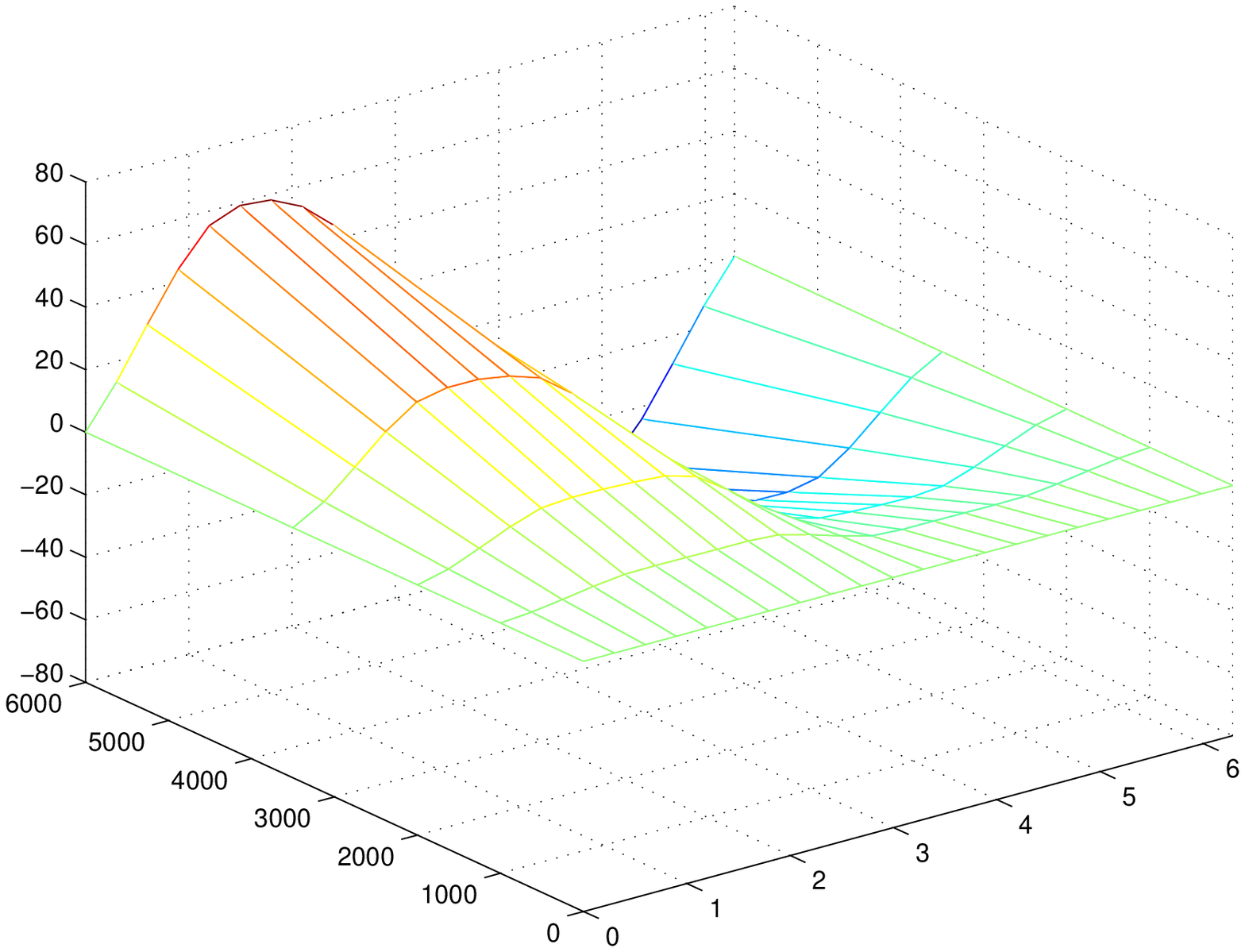}
\else
\includegraphics[width=.4\textwidth]{matlab/mecrow_model/tau.eps}
\includegraphics[width=.4\textwidth]{matlab/tau_hat.eps}
\fi
\caption{
The torque characteristic $g_1$ (left),
and its piecewise-affine approximation $\tilde g_1$ (right).
}
\label{f-torque-function}
\end{center}
\end{figure}

\paragraph{Flux characteristic and phase torque functions.}
The magnetic flux characteristic $f_1$ and the phase torque function $g_1$ 
are shown in figures~\ref{f-flux-characteristic}
and \ref{f-torque-function}.
%We only show the characteristic for positive values of the MMF,
%even though is possible to have negative MMF in a tooth.
We only show these functions for the first magnetic element.
(Due to phase symmetry,
the other two functions are shifted versions of the first.)

For each value of $\theta_t$, for $t = 0, \ldots, T$,
we fit piecewise-affine functions to $f_k$ and $g_k$.
The piecewise-affine region boundaries were chosen as
\[
\tilde {\mathcal F} = (0, 1000, 2000, 3500, 6000) \text{ Ampere-turns}.
\]
With the region boundaries fixed,
the values of the coefficient functions $a_k$, $b_k$, $c_k$, and $d_k$
were determined at the relevant (discrete) rotor positions using least-squares,
with the constraint that both piecewise-affine functions be continuous
in $\mathcal F$ for all values of $\theta$.

%  \begin{figure} 
%  \begin{center}
%  %\psfrag{theta}[cc]{$\theta$} 
%  \includegraphics[width=.4\textwidth]{matlab/phi_hat.eps}
%  \includegraphics[width=.4\textwidth]{matlab/tau_hat.eps}
%  \caption{
%  For $\theta = $,
%  the flux characteristic $f_1$ (solid) and 
%  its piecewise-affine approximation $\tilde f_1$ (dashed)
%  are shown on the left,
%  and the torque function $g_1$ (solid) and 
%  its piecewise-affine approximation $\tilde g_1$ (dashed)
%  are shown on the right.
%  }
%  \label{f-estimated-characteristics}
%  \end{center}
%  \end{figure}

\subsection{Results}

\begin{figure} 
\begin{center}
%\psfrag{theta}[cc]{$\theta$} 
\ifarxiv
\includegraphics[width=.7\textwidth]{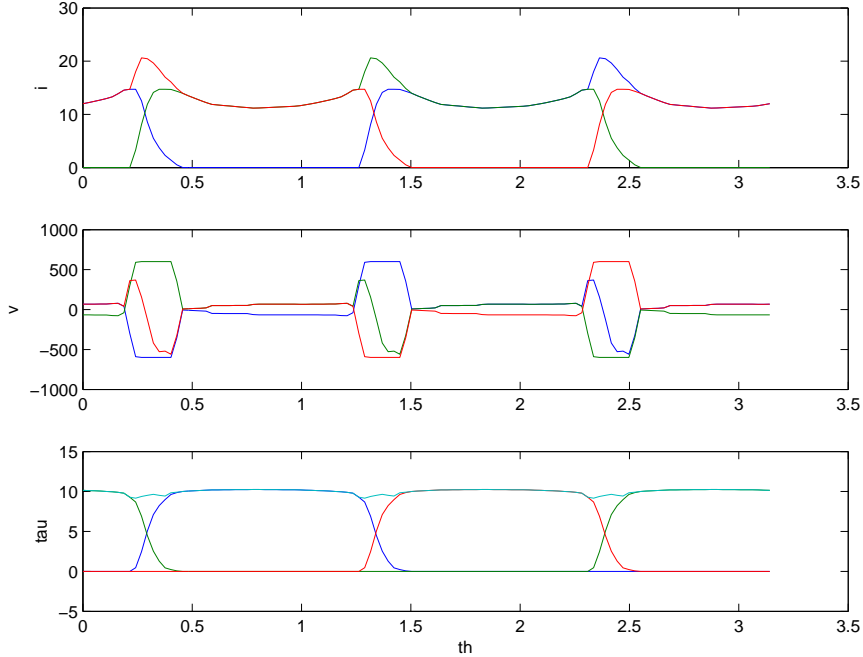}
\else
\includegraphics[width=.7\textwidth]{matlab/waves_low_speed.eps}
\fi
\caption{
The optimal current, voltage, and torque waveforms for
$\omega = 1000$ rpm.
}
\label{f-waves-low-speed}
\end{center}
\end{figure}

\paragraph{Low-speed operation.}
The optimal current, voltage, and phase torque waveforms
for $\omega = 1000$ rpm, $\tau_{\rm des} = 10$ $\rm N\cdot m$ and
$\alpha = 3$ $\rm J/(N\cdot m)^2$
are shown in figure~\ref{f-waves-low-speed}.
(We restricted our search to waveforms that are phase symmetric,
as discussed in $\S\ref{s-symmetry}$.)

Note that while the voltage limit is active over the commutation period
for two of the windings,
the third winding current is manipulated to maintain near-constant torque.
Due to the fully pitched winding pattern,
and in contrast to a motor with concentrated windings, 
it is not possible to derive a ``torque-sharing function'' for this example.
This is because torque cannot be decomposed into components attributable to each winding
(indeed, it is the changing mutual inductance between windings which generates torque;
for a discussion of this, see \cite{mecrow1993fully}).

\paragraph{High-speed operation.}
Here we show the optimal symmetric
current, voltage, and phase torque waveforms
for $\omega = 4000$ rpm,
with all other values kept the same as for the low-speed example.

We see that when the rotor speed is higher, 
the optimal waveforms are much more complicated
than the corresponding low-speed waveforms.
Indeed, for high rotor speeds,
deriving a simple, ``closed-form'' solution
for the optimal current waveforms seems unlikely.
The optimal current waveforms are strictly positive
during the entire cycle;
\ie, there is no ``firing angle''
at which a given phase phase should be energized.

\begin{figure} 
\begin{center}
%\psfrag{theta}[cc]{$\theta$} 
\ifarxiv
\includegraphics[width=.7\textwidth]{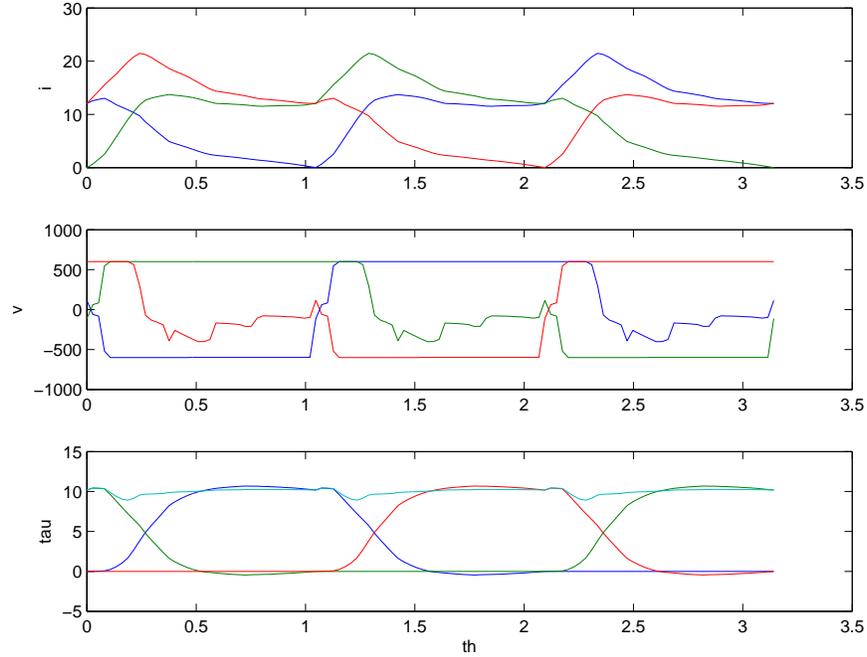}
\else
\includegraphics[width=.7\textwidth]{matlab/waves_hi_speed.eps}
\fi
\caption{
The optimal current, voltage, and torque waveforms for
$\omega = 4000$ $\rm rpm$.
}
\label{f-waves-hi-speed}
\end{center}
\end{figure}

\subsection{Computational aspects}
To compute the waveforms above,
problem (\ref{e-opt-torque-ctrl}) was solved 
using CVX \cite{cvx, gb08},
%with MOSEK \cite{andersen2000mosek} as the backend solver.
with Gurobi as the backend solver.
We used a Linux machine with an Intel Xeon processor.
The (global) solve time for low-speed operation was 6 seconds,
while the solve time for high-speed operation was 4 minutes.

In order to verify that these solve times were representitive
of the solve time over several motor operating points,
we solved (\ref{e-opt-torque-ctrl}), 
using the same motor parameters 
(except for a courser grid of $T = 20$)
for 250 randomly selected values of $\omega$ and $\tau_{\rm des}$.
in the intervals $[0, 4000]$ $\rm rpm$ and $[0, 15]$ $\rm N$, respectively.
The median global solve times was ten seconds,
though we note that for some areas of operation
(in particular, for very-low-torque operation)
the global solve times can be very high, sometimes exceeding thirty minutes;
in these cases, it is standard to terminate the solver early
(after, say, several minutes)
simply use the best point found so far.

\section{Conclusion}
In this paper we presented a method for generating optimal current waveforms
for switched reluctance motors.
Our model handles a reasonably complicated magnetic structure,
and respects voltage and current constraints.
Our method finds, to within reasonable accuracy,
the globally optimal waveforms.
Although we provide to runtime guarentees,
for our example, optimal waveforms could be generated in a few minutes,
raising the possibility that optimal waveforms can be generated
and stored as a lookup table indexed by desired torque and rotor speed.

\newpage
\bibliography{sw_reluc_motor}

\newcommand{\etalchar}[1]{$^{#1}$}
\begin{thebibliography}{TMBB15}

\bibitem[Bal79]{balas1979disjunctive}
E.~Balas.
\newblock Disjunctive programming.
\newblock {\em Annals of Discrete Mathematics}, 5:3--51, 1979.

\bibitem[BV04]{boyd2004convex}
S.~Boyd and L.~Vandenberghe.
\newblock {\em Convex Optimization}.
\newblock Cambridge University Press, 2004.

\bibitem[CKKP02]{choi2002new}
C.~Choi, S.~Kim, Y.~Kim, and K.~Park.
\newblock A new torque control method of a switched reluctance motor using a
  torque-sharing function.
\newblock {\em IEEE Transactions on Magnetics}, 38(5):3288--3290, 2002.

\bibitem[CMH93]{le1993current}
J.~Y.~Le Chenadec, B.~Multon, and S.~Hassine.
\newblock Current feeding of switched reluctance motor. optimization of the
  current waveform to minimize the torque ripple.
\newblock {\em Proceedings of the 4th International Conference on Computational
  Aspects of Electromechanical Energy Converters and Drives}, 1993.

\bibitem[CS99]{ceria1999convex}
S.~Ceria and J.~Soares.
\newblock Convex programming for disjunctive convex optimization.
\newblock {\em Mathematical Programming}, 86(3):595--614, 1999.

\bibitem[CS02]{chapman2002design}
P.~L. Chapman and S.~D. Sudhoff.
\newblock Design and precise realization of optimized current waveforms for an
  8/6 switched reluctance drive.
\newblock {\em IEEE Transactions of Power Electronics}, 17(1):76--83, 2002.

\bibitem[DK84]{desoer1984basic}
C.~A. Desoer and E.~S. Kuh.
\newblock {\em Basic circuit theory}.
\newblock Tata McGraw-Hill Education, 1984.

\bibitem[FG07]{frangioni2007sdp}
A.~Frangioni and C.~Gentile.
\newblock {SDP} diagonalizations and perspective cuts for a class of
  nonseparable {MIQP}.
\newblock {\em Operations Research Letters}, 35(2):181--185, 2007.

\bibitem[GB08]{gb08}
Michael Grant and Stephen Boyd.
\newblock Graph implementations for nonsmooth convex programs.
\newblock In V.~Blondel, S.~Boyd, and H.~Kimura, editors, {\em Recent Advances
  in Learning and Control}, Lecture Notes in Control and Information Sciences,
  pages 95--110. Springer-Verlag Limited, 2008.

\bibitem[GB14]{cvx}
Michael Grant and Stephen Boyd.
\newblock {CVX}: Matlab software for disciplined convex programming, version
  2.1.
\newblock \url{http://cvxr.com/cvx}, March 2014.

\bibitem[GL12]{gunluk2012perspective}
O.~G{\"u}nl{\"u}k and J.~Linderoth.
\newblock Perspective reformulation and applications.
\newblock In {\em Mixed Integer Nonlinear Programming}, pages 61--89. Springer,
  2012.

\bibitem[GT13]{grossmann2013systematic}
I.~E. Grossmann and F.~Trespalacios.
\newblock Systematic modeling of discrete-continuous optimization models
  through generalized disjunctive programming.
\newblock {\em American Institute of Chemical Engineers Journal},
  59(9):3276--3295, 2013.

\bibitem[KSAS03]{kaiserseder2003current}
M.~Kaiserseder, J.~Schmid, W.~Amrhein, and V.~Scheef.
\newblock Current shapes leading to positive effects on acoustic noise of
  switched reluctance drives.
\newblock {\em International Journal for Computation and Mathematics in
  Electrical and Electronic Engineering}, 22(4):998--1008, 2003.

\bibitem[KT06]{kleinberg2006algorithm}
J.~Kleinberg and E.~Tardos.
\newblock {\em Algorithm design}.
\newblock Pearson Education, 2006.

\bibitem[LS94]{lovatt1994computer}
H.~C. Lovatt and J.~M. Stephenson.
\newblock Computer-optimised current waveforms for switched-reluctance motors.
\newblock {\em IEE Proceedings Electric Power Applications}, 141(2):45--51,
  1994.

\bibitem[LS97a]{lovatt1997computer}
H.~C. Lovatt and J.~M. Stephenson.
\newblock Computer-optimised smooth-torque current waveforms for
  switched-reluctance motors.
\newblock {\em IEE Proceedings Electric Power Applications}, 144(5):310--316,
  1997.

\bibitem[LS97b]{lovatt1997optimum}
H.~C. Lovatt and J.~M. Stephenson.
\newblock Optimum excitation of switched reluctance motors.
\newblock In {\em Eighth International Conference on Electrical Machines and
  Drives}, pages 356--360. IET, 1997.

\bibitem[MB15]{moehle2015optimal}
N.~Moehle and S.~Boyd.
\newblock Optimal current waveforms for brushless permanent magnet motors.
\newblock {\em International Journal of Control}, 88(7):1389--1399, 2015.

\bibitem[Mec93]{mecrow1993fully}
B.~C. Mecrow.
\newblock Fully pitched-winding switched-reluctance and stepping-motor
  arrangements.
\newblock {\em IEEE Proceedings Electric Power Applications}, 140(1):61--70,
  1993.

\bibitem[MK03]{mademlis2003performance}
C.~Mademlis and I.~Kioskeridis.
\newblock Performance optimization in switched reluctance motor drives with
  online commutation angle control.
\newblock {\em IEEE Transactions on Energy Conversion}, 18(3):448--457, 2003.

\bibitem[MWC01]{mecrow2001modeling}
B.~C. Mecrow, C.~Weiner, and A.~C. Clothier.
\newblock The modeling of switched reluctance machines with magnetically
  coupled windings.
\newblock {\em IEEE Transactions on Industry Applications}, 37(6):1675--1683,
  2001.

\bibitem[NW06]{nocedal2006numerical}
J.~Nocedal and S.~Wright.
\newblock {\em Numerical optimization}.
\newblock Springer Science \& Business Media, 2006.

\bibitem[TK12]{toliyat2012handbook}
H.~A. Toliyat and G.~B. Kliman.
\newblock {\em Handbook of electric motors}, volume 120.
\newblock CRC press, 2012.

\bibitem[TMBB15]{takapoui2015simple}
R.~Takapoui, N.~Moehle, S.~Boyd, and A.~Bemporad.
\newblock A simple effective heuristic for embedded mixed-integer quadratic
  programming.
\newblock {\em arXiv preprint arXiv:1509.08416}, 2015.

\bibitem[VZP{\etalchar{+}}07]{vasak2007bounding}
M.~Vasak, D.~Zarko, N.~Peric, F.~Kolonic, and C.~Hao.
\newblock Bounding the torque ripple in switched reluctance motors using
  polyhedral invariant set theory.
\newblock In {\em 46th IEEE Conference on Decision and Control}, pages
  6106--6111. IEEE, 2007.

\end{thebibliography}

\end{document}